\documentclass{aa}  
\usepackage{graphicx}
\usepackage{txfonts}
\begin{document} 

   \title{Investigating the Long Secondary Period Phenomenon\\ with the ASAS-SN and Gaia data}
   \titlerunning{LSP in ASAS-SN and Gaia}

   \author{Micha{\l} Pawlak$^{1,2}$, Michele Trabucchi$^{3,4}$, Laurent Eyer$^{4}$, and Nami Mowlavi$^{4,5}$}
          
    \authorrunning{M. Pawlak et al.}

   \institute{$^{1}$ Astronomical Observatory, Jagiellonian University, ul. Orla 171, 30-244 Krak{\'o}w, Poland\\
   $^{2}$ Lund Observatory, Division of Astrophysics, Department of Physics, Lund University, Box 43, SE-221 00, Lund, Sweden\\
   $^{3}$ Dipartimento di Fisica e Astronomia Galileo Galilei, Università degli studi di Padova, Vicolo dell’Osservatorio 3, I-35122 Padova, Italy\\
   $^{4}$ Department of Astronomy, University of Geneva, Chemin Pegasi 51, 1290, Versoix, Switzerland\\
   $^{5}$ Department of Astronomy, University of Geneva, Chemin d'Ecogia 16, 1290, Versoix, Switzerland\\
             \\ \email{michal.pawlak@fysik.lu.se}
             }
   \date{}

 
  \abstract
   {}
   {The aim of this work is to create a complete list of sources exhibiting a  long secondary period (LSP) in the ASAS-SN catalog of variable stars, and analyze the properties of this sample compared to other long period variables without LSP.}
   {We use the period-amplitude diagram to identify the 55\,572 stars showing an LSP, corresponding to 27\% of the pulsating red giants in the catalog. We use the astrometric data from {\it Gaia} DR3 and the spectroscopic data provided by the APOGEE, GALAH, and RAVE surveys to investigate the statistical properties of the sample.}
   {We find that stars displaying an LSP have a spatial distribution that is more dispersed than the non-LSP giants, suggesting that they belong to an older population. Spectroscopically-derived ages seem to confirm this. The stars with an LSP also appear to be different in terms of C/O ratio from their non-LSP counterparts.}
   {}


   \maketitle
%

\section{Introduction}

The long secondary period (LSP) phenomenon is observed in a significant fraction of long period variables (LPVs). It is an additional source of periodic variability which exists alongside the primary, pulsational variability but cannot be explained by radial pulsation. Despite the fact that the phenomenon has been known for a long time \citep{oconnell1933, paynegaposchkin1954, houk1963}, its origin still lacks a full explanation. A number of hypotheses have been put forward to explain the origin of LSP. The two most common include binarity \citep{wood1999,soszynski2007,soszynski2014,soszynski2021} and non-radial pulsation \citep{wood2000a,wood2000b,hinkle2002,wood2004,saio2015}.

LSP stars were spectroscopically observed by \citet{nicholls2009}, who measured their radial velocities in order to verify the binary hypothesis, concluding that the classical binary with a stellar companion can be ruled out. However, neither the low-mass companion scenario nor the one involving non-radial pulsations could be excluded. An analysis of the spectroscopic properties of a small sample of LSPs was carried out by \citet{pawlak2019}, who noted some possible differences in basic spectroscopic parameters including $\log g$, effective temperate and metallicity. A similar study was done by \citet{jayasinghe2021}, who concluded that no significant difference can be seen.

In this paper, we use the All Sky Automated Survey for Supernovae \citep[ASAS-SN]{shappee2014,jayasinghe2018} catalog of LPVs combined with {\it Gaia} Data Release~3 \citep[DR3]{GaiaEDR3,GaiaDR3} to identify a complete, all-sky sample of LSPs and study their properties compared to the non-LSP red giants. We further extend our analysis using spectroscopic data provided by the APOGEE \citep{majewski2017}, Galah \citep{desilva2015}, and Rave \citep{steinmetz2006} surveys. 

The structure of the paper is as follows: Sect.~2 describes the data set we used and Sect.~3 presents the analysis of the spatial distribution, kinematics as well as spectroscopic properties of the sample. We discuss the results in Sect.~4.  

\section{Data}

For the purpose of this study we use the ASAS-SN Catalog of Variable Stars containing $194\,840$ LPVs \citep{jayasinghe2018, jayasinghe2019,jayasinghe2019b}. The sample includes both Semi-Regular as well as OGLE Small-Amplitude Red Giant (OSARG) type variables. We cross-match this catalog with the {\it Gaia} DR3 catalog \citep{GaiaEDR3}, obtaining $186\,583$ matches. This is our base sample which we use for further analysis. 

\begin{figure*}
   \centering
       \includegraphics[width=0.75\textwidth]{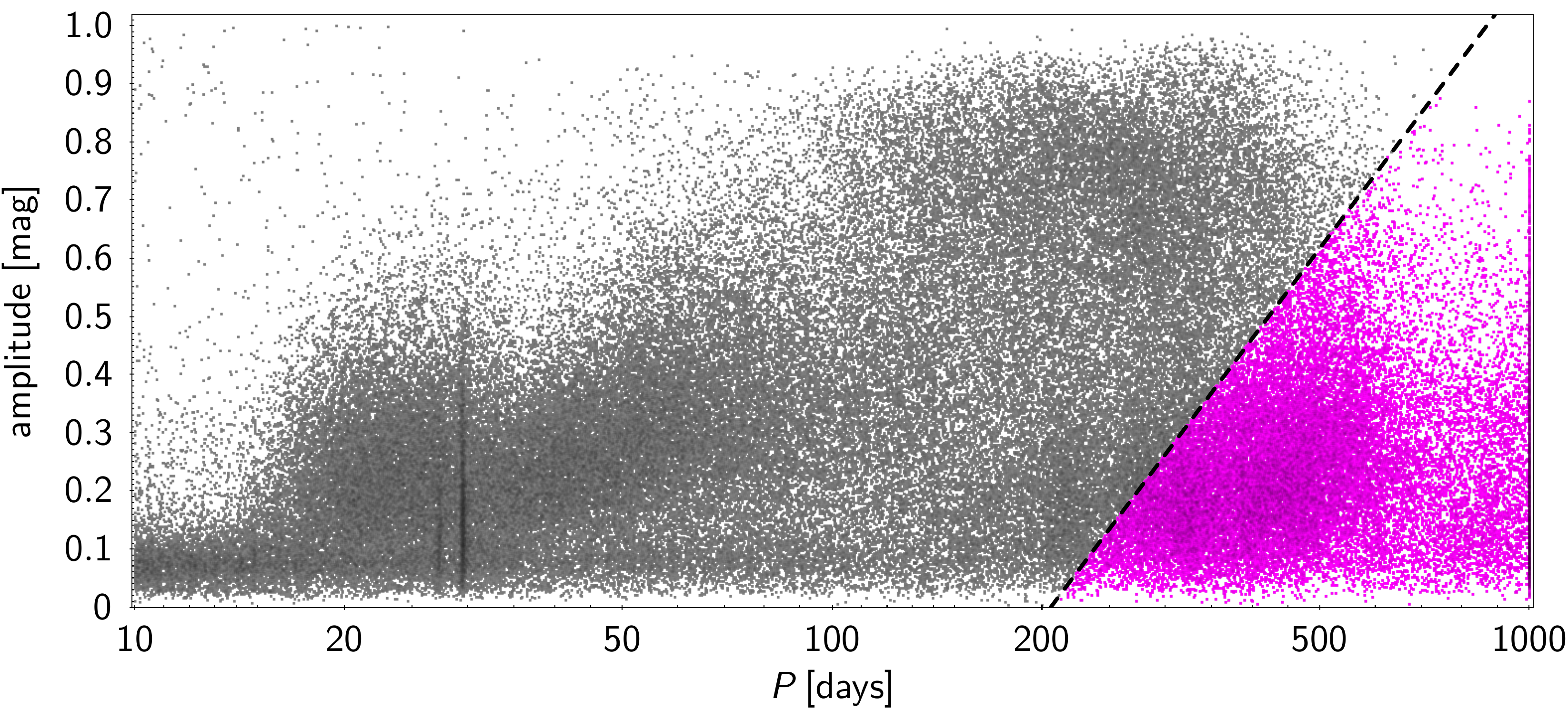} 
 \caption{Selection of the LSP stars based on the location on the period-amplitude diagram. Periods selected as LSPs are marked in magenta, while other LPVs are marked in grey. The cut-off criterion is $A < 1.6\log(P)-3.7$, and it is marked with a black, dashed line.}
 \label{fig:lsp_sel}
\end{figure*}

\begin{figure*}
   \centering
       \includegraphics[width=0.75\textwidth]{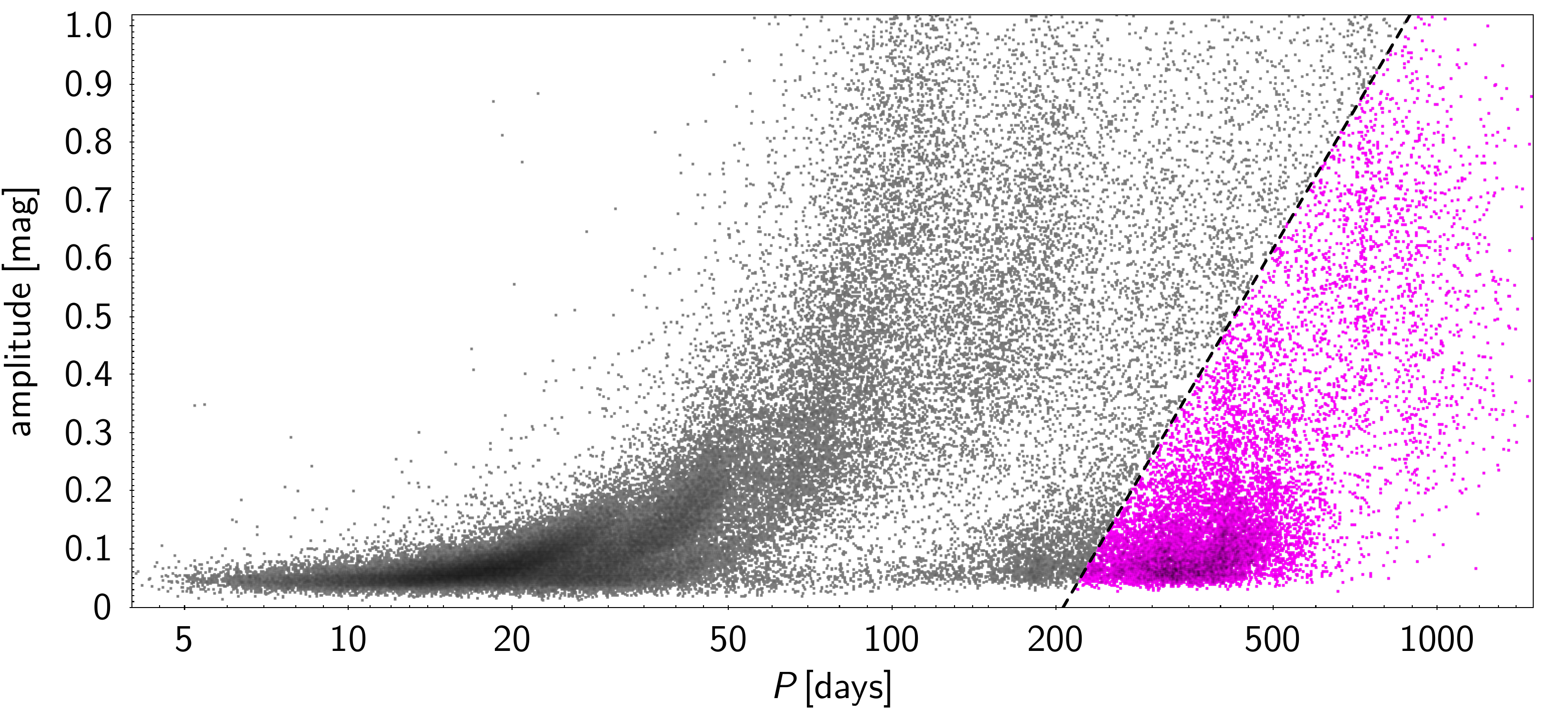}
     \includegraphics[width=0.76\textwidth]{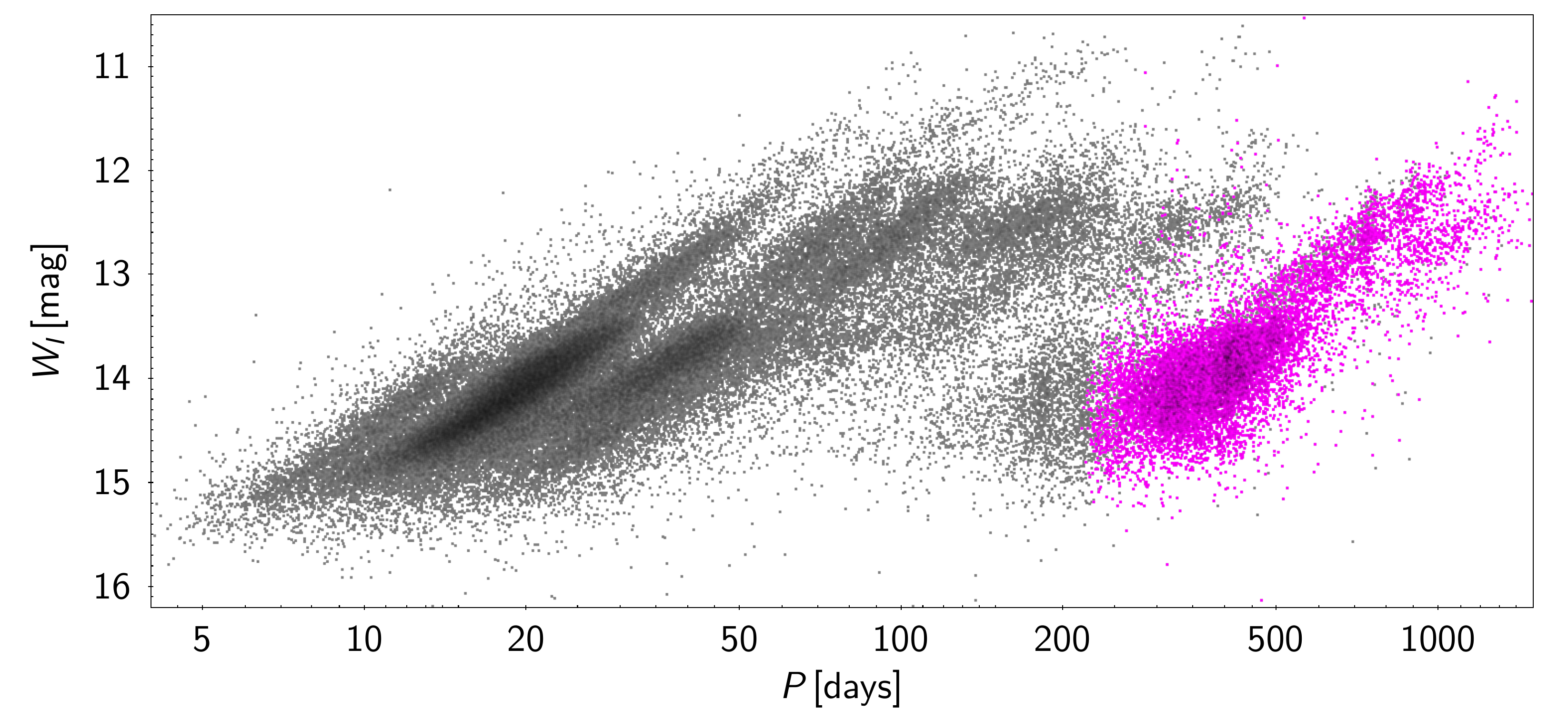}
 \caption{Verification of the selection criterion using the OGLE LMC sample \citep{soszynskietal2007}. The period-amplitude cut-off marked with dashed line is the same as in Fig.~1. The objects selected with the cut-off (upper panel) are placed on the PL diagram (lower panel). For the PL diagram, we use the redenig-free Wesenheit index $W_I = V -1.55(V-I)$.}
 \label{fig:lsp_sel_ogle}
\end{figure*}

\begin{figure*}
   \centering
       \includegraphics[width=0.75\textwidth]{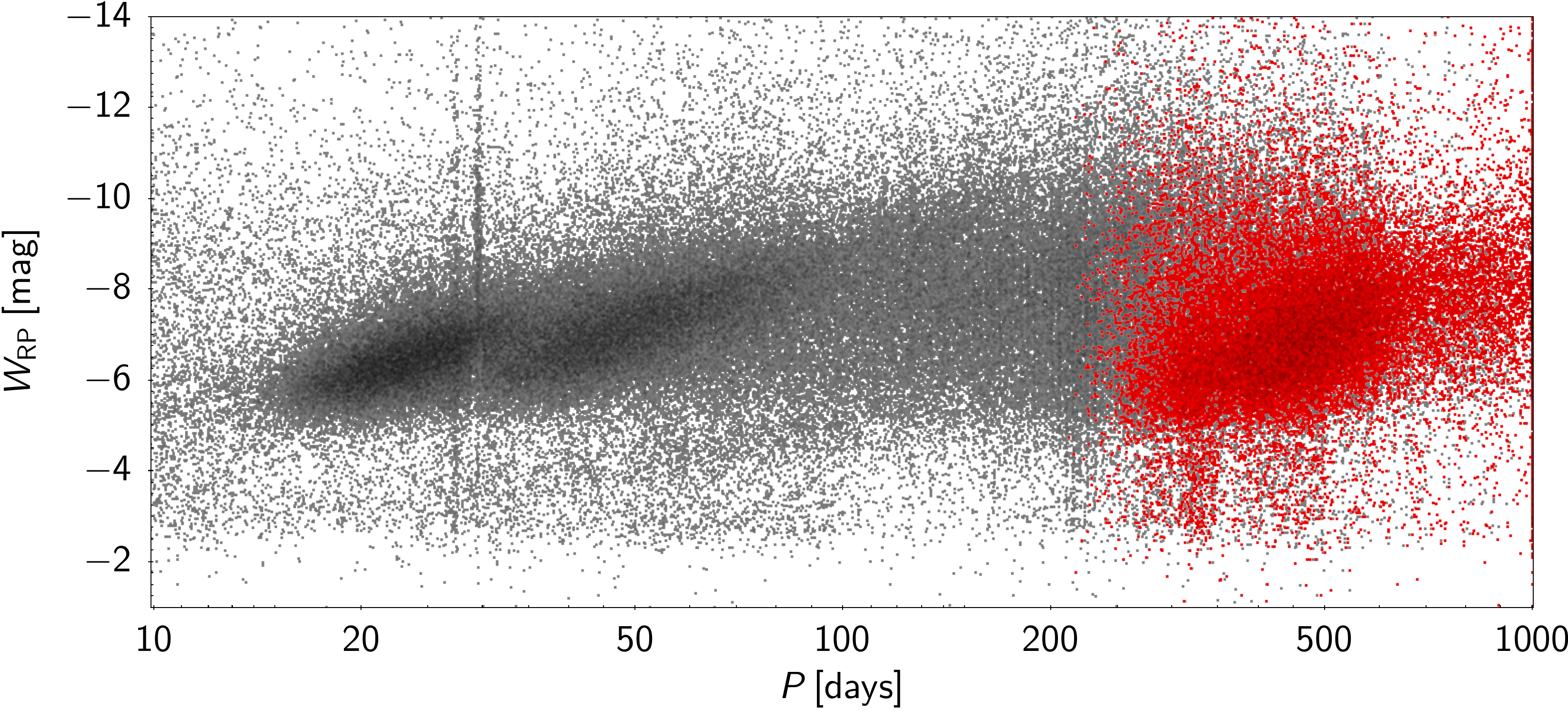} \\
       \includegraphics[width=0.75\textwidth]{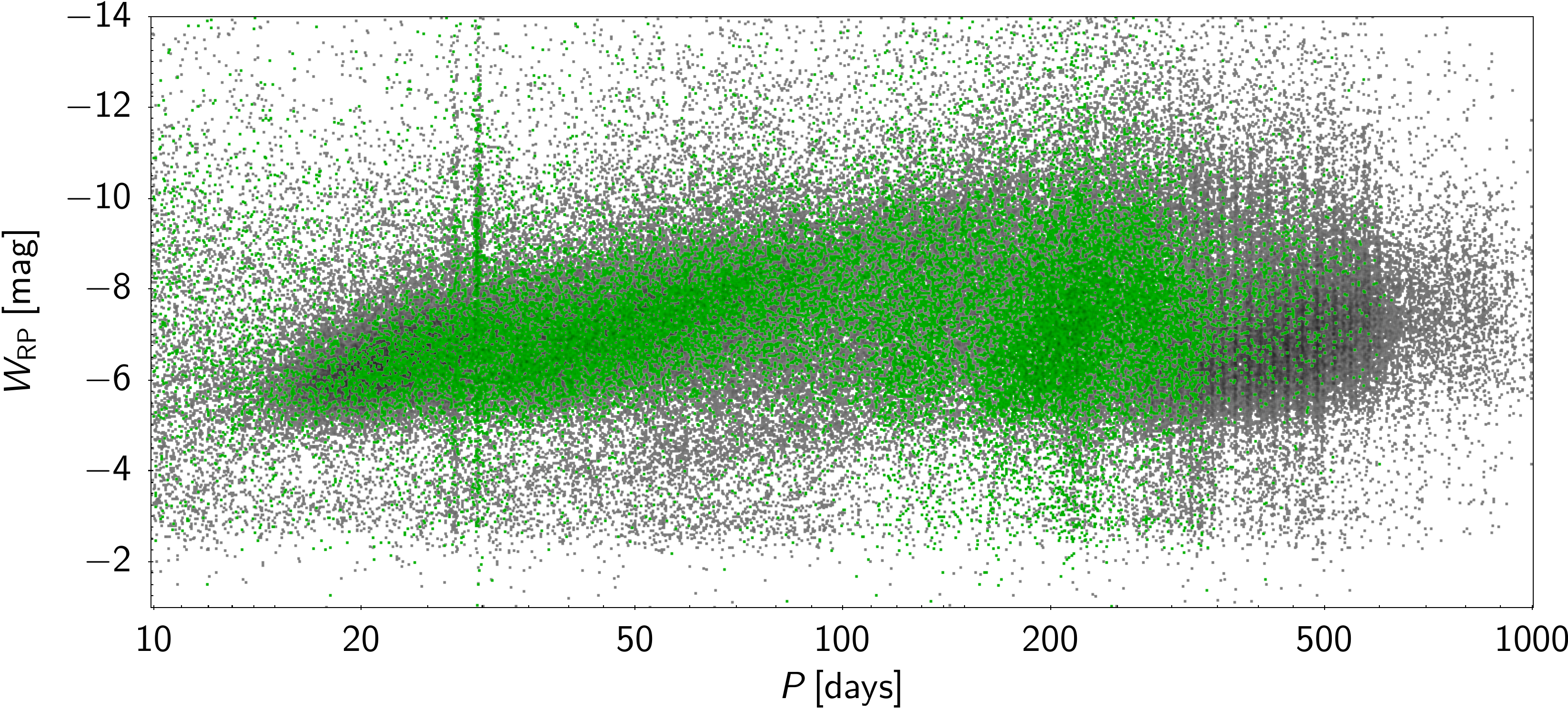} 
 \caption{The location of the stars that have LSP as the strongest period in the period vs. {\it Gaia} Wesenheit index plane. In the top panel the stars form the sample are plotted using the LSPs as the period (red). In the bottom-panel the same stars are plotted with the strongest pulsational period (green). Non-LSP LPVs are shown as a background in grey in both panels.} 
 \label{fig:lsp_1}
\end{figure*}

The dominant variability period is given in the ASAS-SN catalog. However, since LSP can appear not only as the strongest but also as the second or further period, we decide to run an independent period search to identify the three strongest periods for each of the objects. For that purpose, we use the Lomb-Scargle method implemented in the VAROOLS package \citep{hartman2016}.

\begin{figure*}
   \centering
       \includegraphics[width=0.75\textwidth]{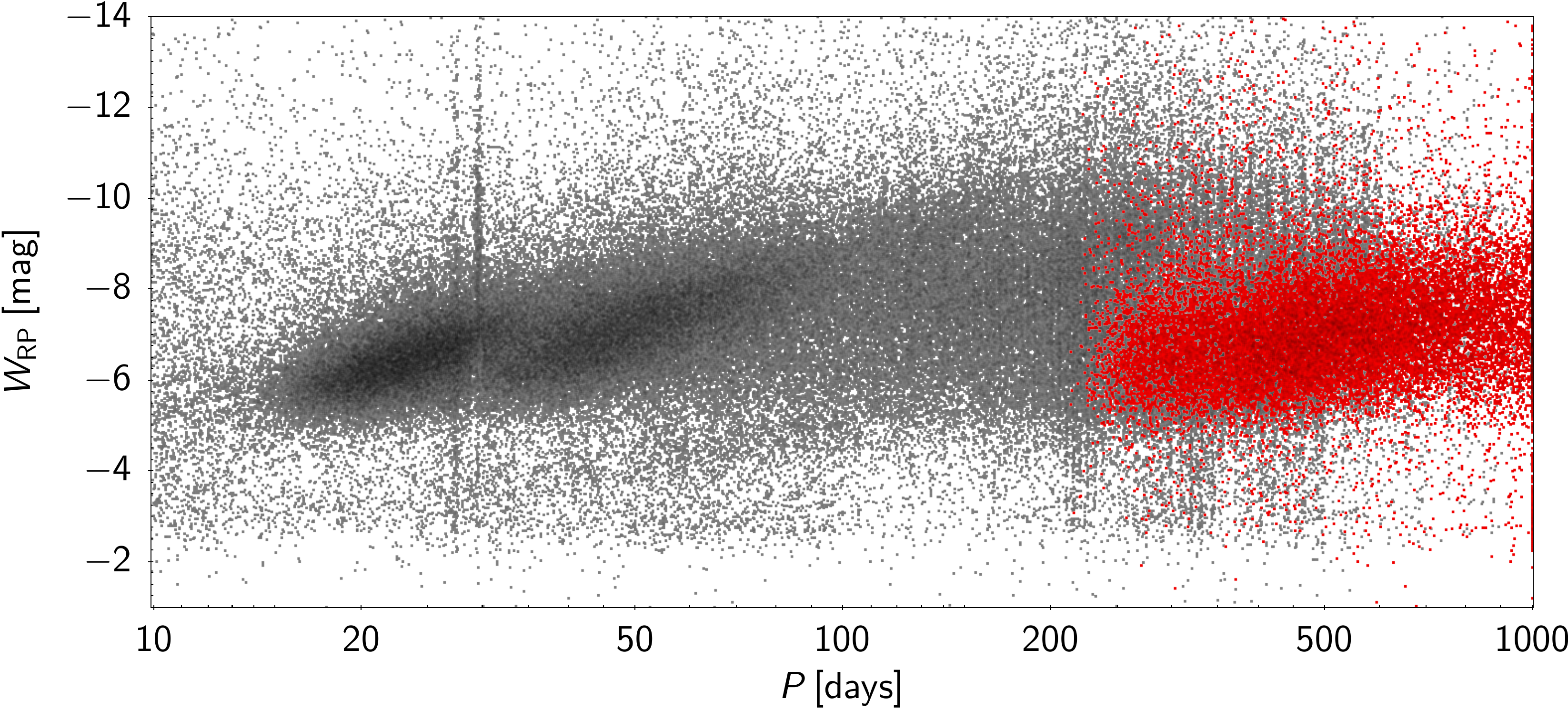} \\
       \includegraphics[width=0.75\textwidth]{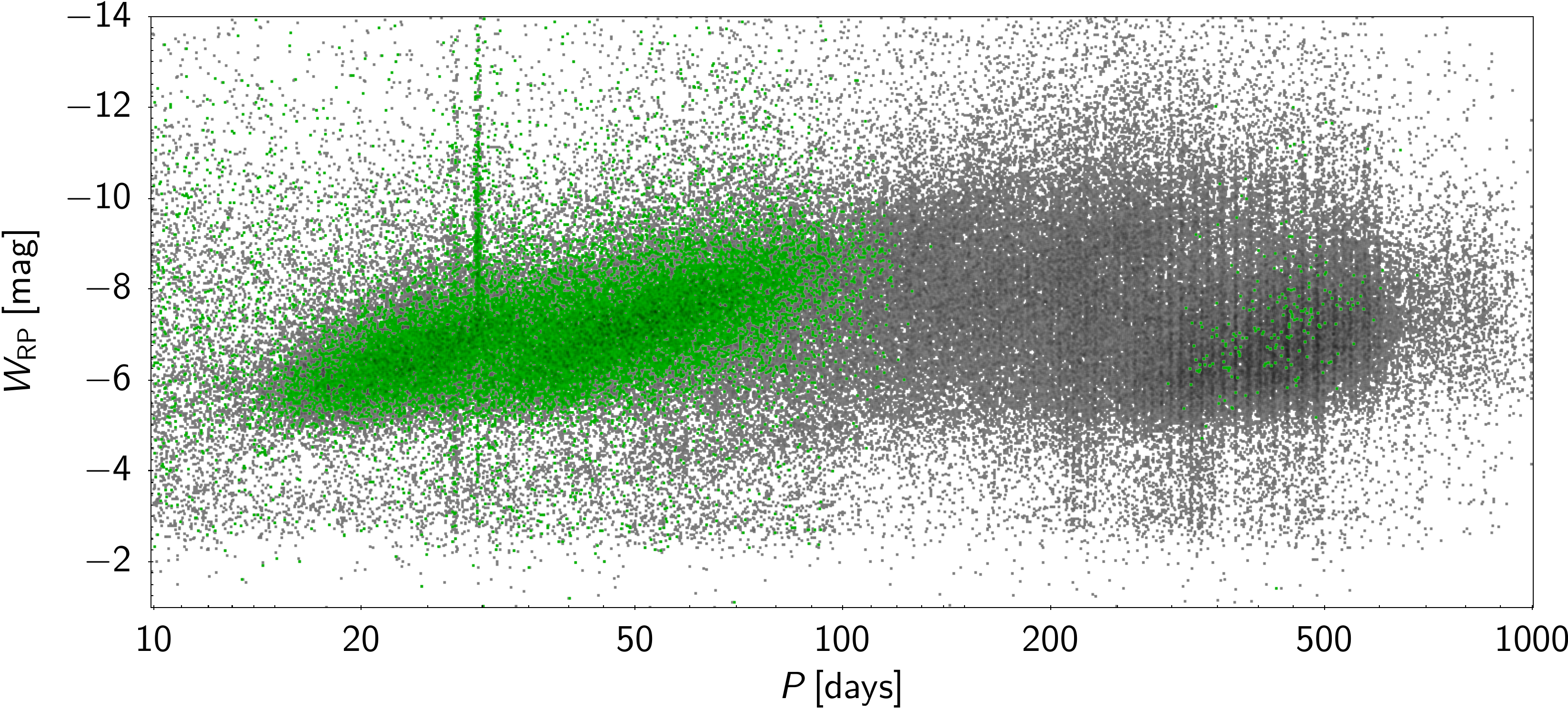} 
 \caption{The same as Fig.~3, but for the stars that have LSP as a 2nd or 3rd period.} 
 \label{fig:lsp_23}
\end{figure*}

\begin{figure*}
   \centering
       \includegraphics[width=0.85\textwidth]{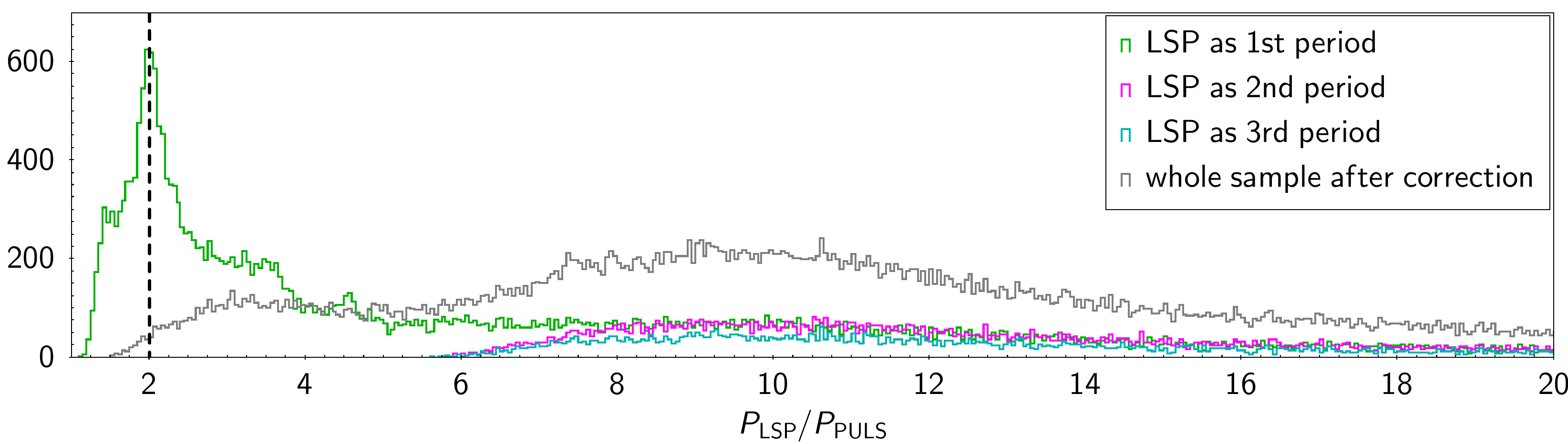} 
 \caption{Period ratio for the stars where LSP is detected as 1st period (green), 2nd period (magenta), and 3rd period (blue). The high pick at $P_{LSP}/P_{PULS} = 2$ is a clear indication that some of the periods identified as pulsational are actually harmonics of the LSP. The whole sample after alias correction is shown in grey.} 
 \label{fig:lsp_ratio}
\end{figure*}

\begin{figure*}
   \centering
       \includegraphics[width=0.75\textwidth]{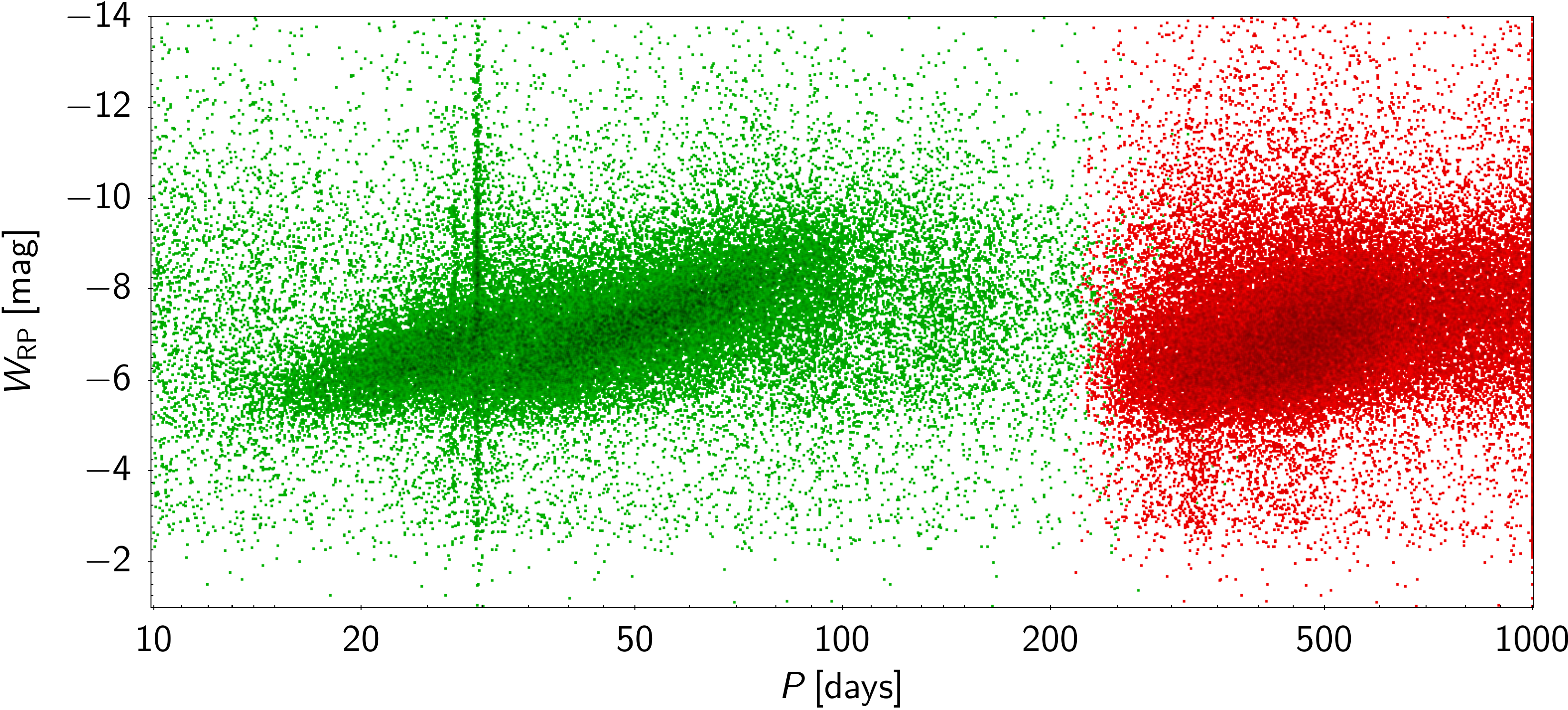} 
 \caption{PL after correction, the color schema is the same as in Fig.~\ref{fig:lsp_1}.  For the stars that have LSP detected as the 1st period and LSP alias as the 2nd period, the 3rd strongest period is adopted as the pulsational period. }
 \label{fig:lsp_sel_final}
\end{figure*}

The cut we use to select LSP stars is: $A < 1.6\log(P)-3.7$, where $P$ is the strongest period in days and $A$ is the $V$-band peak-to-peak amplitude in magnitudes. The selection criterion we propose has been empirically chosen with a trial and error approach to optimally separate the LSP stars from long-period pulsators. Fig.~\ref{fig:lsp_sel} illustrates the selection method.  We first check the period with the strongest $S/N$ and if it does not meet the LSP criterion, we check the second and third periods.

The LSP stars have been traditionally selected based on their location on the PL diagram. This approach has certain limitations related to the fact, that it requires accurate distances and mean NIR or MIR photometry. In the absence of accurate distance determinations, the sequences C (formed by the fundamental mode pulsators) and D (formed by the LSP stars) tend to overlap, making the distinction between Mira periods and LSPs rather difficult. Interstellar extinction further complicates the issue.

The method we propose is meant to achieve the same result avoiding the aforementioned limitations. For that purpose period-amplitude selection can be used as an effective tracer of the sequence D \citep{trabucchi2017, lebzelter2019}.

To test the reliability of our method we use the OGLE catalogue of LPV stars in the Large Magellanic Cloud \citep{soszynski2007}, where the PL sequences are well-defined. We select the LSP candidates using the period-amplitude criterion defined above and place them on the PL diagram (Fig.~\ref{fig:lsp_sel_ogle}). The vast majority of the selected candidates lie on sequence D, with very few outliers, showing that the period-amplitude cut is an effective way of selecting LSP stars. It also allows us to separate the LSP stars on the sequence D from the ellipsoidal binaries on the tip of the sequence E (the clump of stars with periods between 100 and 200 days, to the left form the LSP sequence.)

We note that our ability to detect LSP is dependent on the magnitude of the star, as the root mean square (RMS) of ASAS-SN photometry gets higher for fainter objects. As shown in \citet{jayasinghe2019b}, the typical RMS remains at the level of 0.01~mag up to $V = 13$~mag and then rises steadily to about 0.1~mag at $V =17$~mag. Therefore, to avoid spurious detection we introduce a cut on amplitude for the faint stars, defined as $A > 0.036 \cdot V -0.458$. In the whole sample, we identify $32\,983$ stars where LSP is the strongest period and $22\,831$ where LSP is the second or third strongest period.

In order to verify the performance of the selection method proposed as well as further analyze the period-luminosity relations, we compute the Wesenheit reddening-free indexes, defined as in \citet{lebzelter2019}, using both {\it Gaia} and 2MASS photometry and correct it for the distance module based on {\it Gaia} parallaxes. We discard the stars with negative parallaxes. The formulas for {\it Gaia} ($W_{RP}$) and 2MASS ($W_{JK}$) indexes are given by:
\begin{equation}
W_{RP} = G_{RP} - 1.3(G_{BP}-G_{RP})
-5\log(1/\varpi) +5    
\end{equation}
\begin{equation}
W_{JK} = J - 0.686(J-K)-5\log(1/\varpi) +5    
\end{equation}
where, $\varpi$ is the {\it Gaia} parallax, $G_{BP}$, $G_{RP}$ are {\it Gaia} blue and red photometer magnitudes, and $J$, $K$ are 2MASS infrared magnitudes. In Fig.~\ref{fig:lsp_1} we display the period-luminosity diagram (PLD) of stars in which the LSP is dominant, showing both the LSP (top panel) and the pulsation periods (bottom panel). The same diagram is shown in Fig.~\ref{fig:lsp_23} for the stars whose strongest period is not an LSP, but is rather due to pulsation. The first thing we can observe is that the periods that we flagged as LSP, based on period-amplitude are actually located on the sequence D, showing that our selection method is consistent with the traditionally used method based on location on PLD.
We note that the two types of star, shown in Fig.~\ref{fig:lsp_1} and Fig.~\ref{fig:lsp_23}, display clearly different distributions. When the LSP is weaker than pulsation periods, the latter tend to be shorter than 100 days, and populate sequences A and B (for the sequence labels we adopt the same nomenclature as Wood (2015)). Conversely, the pulsation periods of stars dominated by the LSP can be longer than 100 days, and reach the area associated with sequences C', F, and C. 

The comparison of the LSP to pulsation period ratio (Fig.~\ref{fig:lsp_ratio}) reveals the likely cause of the difference mentioned above. For the objects with LSP discovered as the second or third period, the $P_{LSP}/P_{PULS}$ ratio does not show any preferential value. However, for stars where LSP is the strongest period the distribution has a very prominent peak at $P_{LSP}/P_{PULS} = 2$. It strongly suggests that in these cases the second period is not an actual pulsational period, but a harmonic of the LSP. Therefore, for the objects where the second period is longer than 100~d, we take the third period and adopt it as $P_{PULS}$. We also identify 241 where the putative LSP turned out to be an alias of the high-amplitude, long pulsational period. We remove those from our LSP list. The sample corrected this way is shown in Fig.~\ref{fig:lsp_sel_final}.

We make the list of identified LSP, together with all three identified periods, including their pulsational and long periods publicly available in electronic format at CDS\footnote{via anonymous ftp to \url{cdsarc.u-strasbg.fr} (130.79.128.5) or via \url{http://cdsweb.u-strasbg.fr/cgi-bin/qcat?J/A+A/}}.

\section{Analysis}

\subsection{Comparison Sample}

We  further examine our sample of LSP stars in order to assess if they stand out from other LPVs in terms of population effects or spectroscopic properties. It is known that the appearance of LSPs is related to the evolutionary status of the star \citep{trabucchi2017,pawlak2021}. The least-evolved LPVs do not show LSP at all, and the more evolved on the RGB or AGB the star gets, the more likely it is to show an LSP. This effect needs to be taken into account when comparing the LSP and non-LSP samples, as it will obviously bias the analysis. Therefore, to make a proper comparison, we need to define a control sample of non-LSP stars, that are approximately in the same evolutionary point as the LSP sample stars.   
For that purpose, we use the $\log(P)$ vs. $W_{JK}$ plane, where for LSP stars we use the pulsational period as $P$. We remove the stars with negative parallaxes and those with $\varpi <{\sigma}_{\varpi}$, where ${\sigma}_{\varpi}$ is the {\it Gaia} parallax uncertainty. We define X and Y coordinates by normalizing both the $\log(P)$ and $W_{JK}$ range to [0:1]. This way, we obtain a rectangular XY coordinates grid. Then, we take each LSP star and use the simple 2D Cartesian metric to find the nearest non-LSP star. This way we end up with an equally-numerous control sample of non-LSP stars that have approximately the same distribution in the PL diagram as the sample of LSP-dominated stars. In other words, we obtain a control sample of stars that could have shown LSP but do not.

To make sure that the construction of the control sample does not introduce any additional bias, we construct another, randomized control sample. In this case, instead of using the exact $W_{JK}$, we take a value randomly drown from the [$W_{JK}-{\sigma}_{W_{JK}}$, $W_{JK}+{\sigma}_{W_{JK}}$] interval and use it to compute the $Y$ coordinate. Since we consider the uncertainty of $P$ negligible, we leave the $X$ coordinate unchanged. We obtain the randomized coordinates for each of the LSP stars and then select a new nearest non-LSP star. The additional, randomized control sample obtained this way can be used for the purpose of performing quality checks. In the following, the sets constructed with this approach will be indicated as the control sample and the randomized control sample, while we will refer to the set of stars that do or do not show an LSP, as LSP sample and non-LSP sample, respectively.

\subsection{Spatial Distribution and Dynamical properties}

\begin{figure*}
   \centering
       \includegraphics[width=\textwidth]{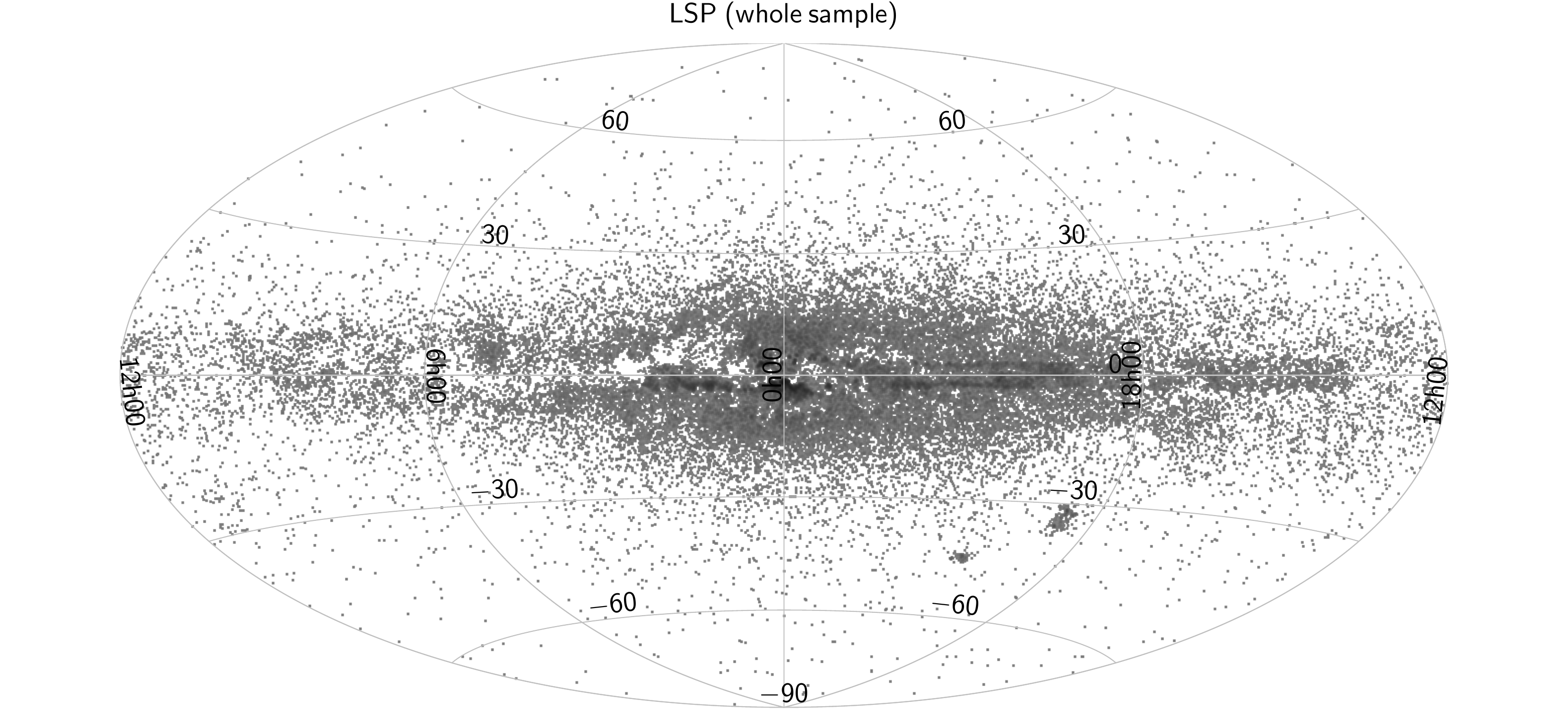} \\ \bigskip \bigskip \bigskip
       \includegraphics[width=\textwidth]{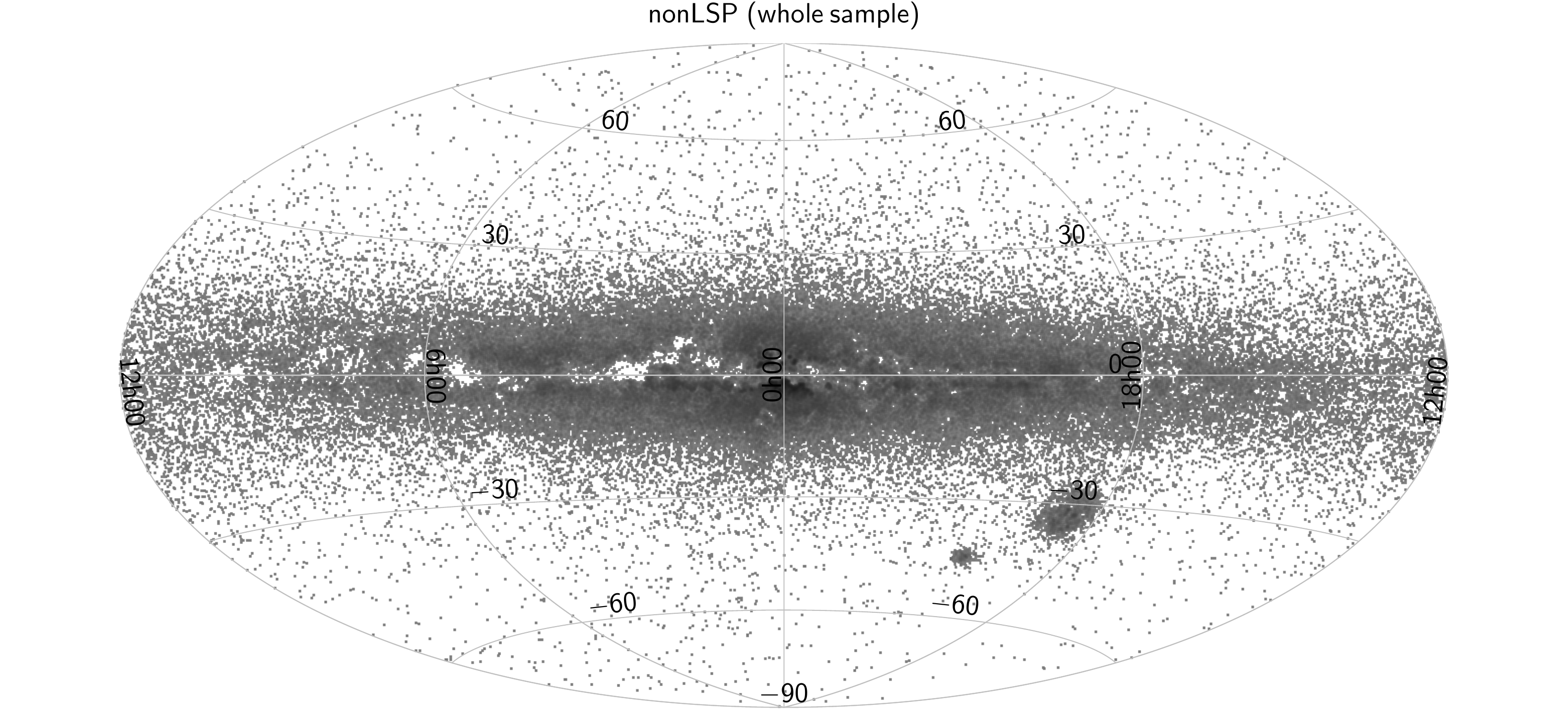} 
 \caption{The sky distribution of the two selected samples. The LSP sample is shown in the top panel and the non-LSP sample in the bottom panel.} 
 \label{fig:sky_all}
\end{figure*}

\begin{figure*}[t]
   \centering
       \includegraphics[width=0.85\textwidth]{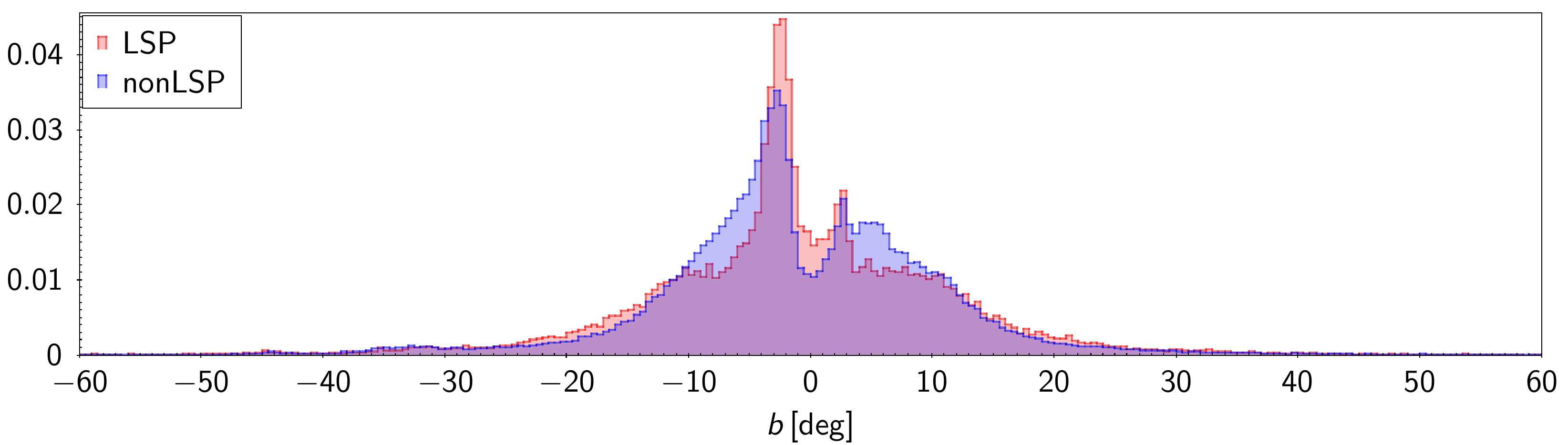} 
 \caption{Galactic latitude distribution for LSP and non-LSP stars. The level of statistical significance tested with the KS test is $p \approx 10^{-16}$.}
 \label{fig:b}
\end{figure*}
\begin{figure*}
   \centering
       \includegraphics[width=0.85\textwidth]{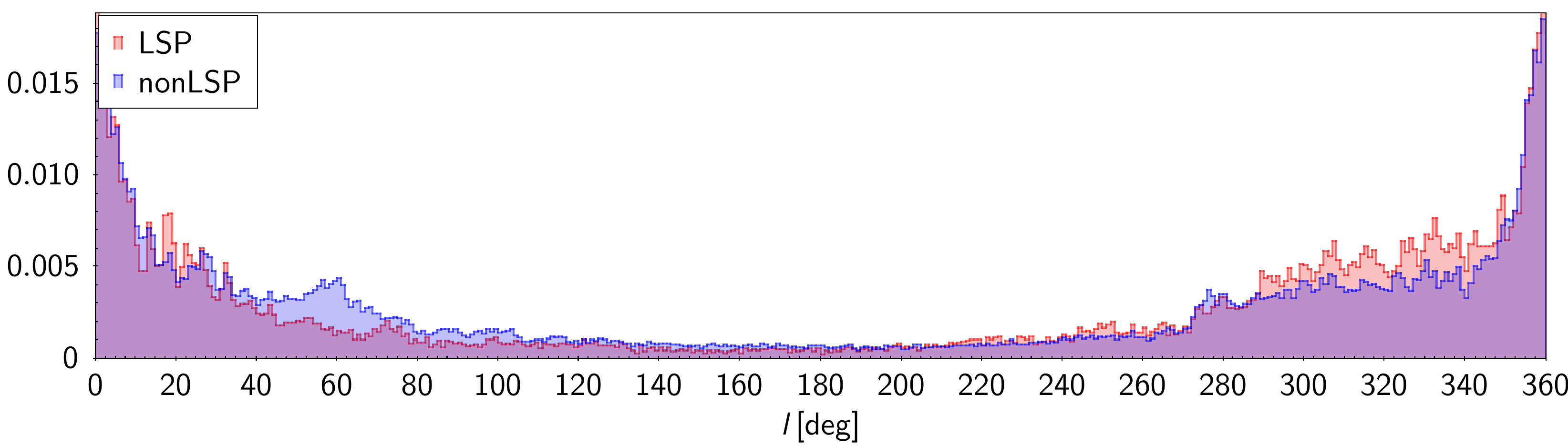} 
 \caption{Galactic longitude distribution. The level of statistical significance tested with the KS test is $p \approx 10^{-16}$.}
 \label{fig:l}
\end{figure*}
\begin{figure*}
   \centering
       \includegraphics[width=0.85\textwidth]{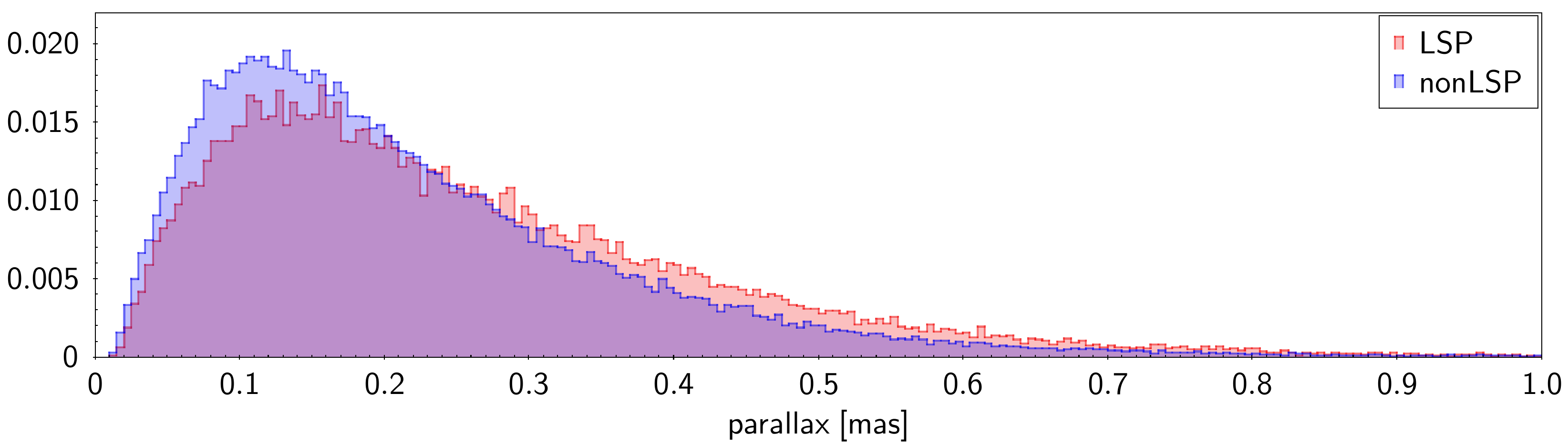} \\
       \includegraphics[width=0.85\textwidth]{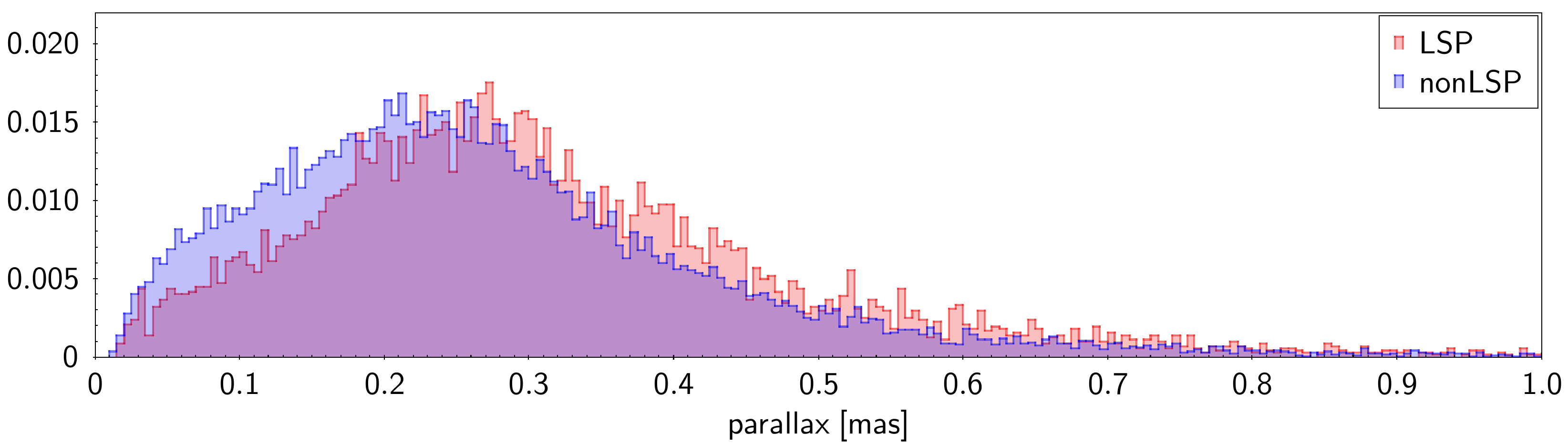}
 \caption{Parallaxes observed towards the Galactic center (upper panel) and the Galactic anti-centre (lower panel).}
 \label{fig:par}
\end{figure*}

We compare the LSP sample to both the whole non-LSP sample and the control sample. First, we examine their sky spatial distribution which is illustrated in Fig.~\ref{fig:sky_all}. We note that the LSP sample seems to be more concentrated around the Galactic bulge and has larger dispersion along the galactic latitude ($b$). 
To check this, we compare the distribution of the samples in $b$. This is illustrated in Fig.~\ref{fig:b}. The distribution of the LSP sample shows a statistically significant difference and indeed appears to be more dispersed in $b$, especially on the negative side of the distribution. To complete the spatial distribution picture, we also compare the distribution in galactic longitude $l$ (Fig.~\ref{fig:l}) and in the  {\it Gaia} parallaxes $\varpi$, which is shown in Fig.~\ref{fig:par} towards both the center and anti-center of the Galaxy. Again, we see a significant difference between LSP and non-LSP stars.

\begin{figure*}
   \centering
       \includegraphics[width=0.85\textwidth]{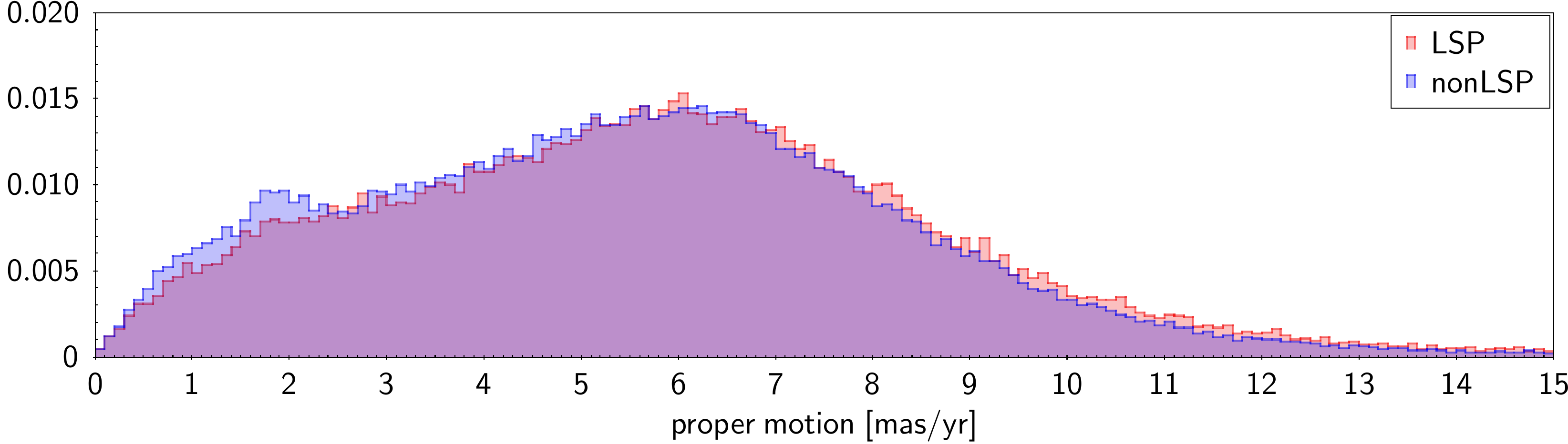} 
 \caption{Proper motion distribution. The level of statistical significance tested with KS test is $p = 6.1\times 10^{-14}$.}
 \label{fig:pm}
\end{figure*}

Next, we investigate the dynamical properties of the sample using {\it Gaia} proper motions. Figure~\ref{fig:pm} shows the distribution in the proper motion of the LSP and non-LSP stars. Again, the distributions are statistically different, with the LSP sample having on average higher proper motions. 

The higher dispersion in $b$ and higher proper motion are both typical of the thick disk population as opposed to the thin disk. The fact that the LSP sample shows both of these features may suggest that stars belonging to the thick disk are more likely to display an LSP compared to stars in the thin disk.

\subsection{Spectroscopic Properties}

\begin{figure*}
   \centering
       \includegraphics[width=0.85\textwidth]{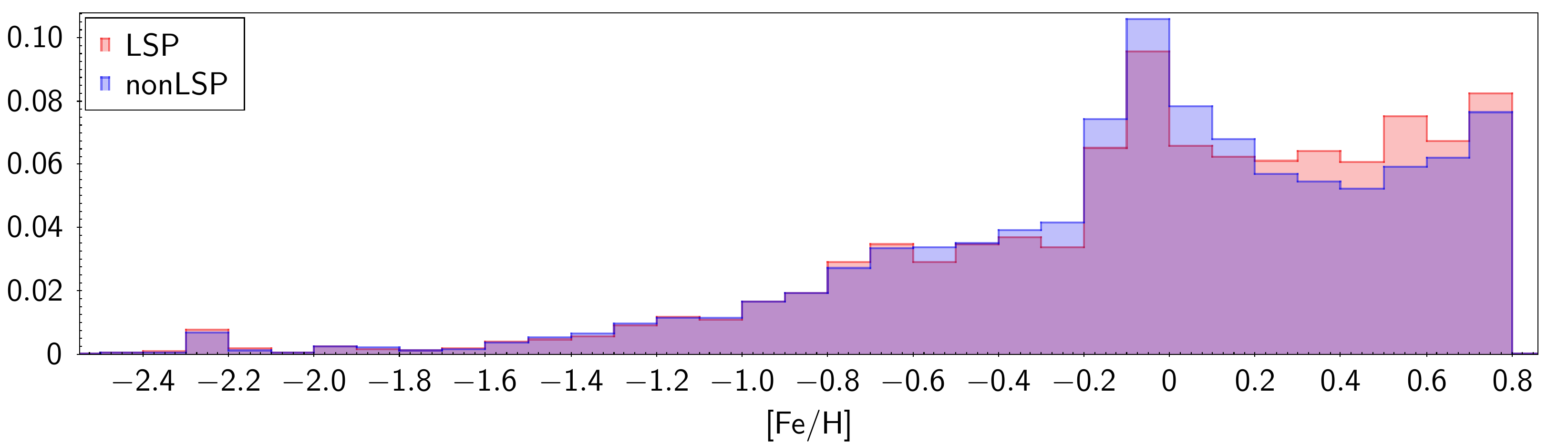} 
 \caption{Metallicilty distribution based on the {\it Gaia} DR3 data. The level of statistical significance tested with KS test is $p \approx 10^{-16}$.}
 \label{fig:feh}
\end{figure*}

For the purpose of further analysis, we cross-match our sample with the following catalogs of spectroscopic surveys: APOGEE DR17 \citep{abdurrouf2022}, GALAH DR3 \citep{buder2021}, and RAVE DR6 \citep{steinmetz2020}. The cross-match results in 652, 916, and 1877 matches for APOGEE, GALAH, and RAVE, respectively. A first analysis of basic spectroscopic parameters, namely ${\log}g$ and $T_{\rm eff}$, did not highlight any clear difference between our samples. We also look into [Fe/H], [M/H], [$\alpha$/H] (if available in a given survey). However, the number of objects for which these measurements are available is too small to make a meaningful comparison. We also check the [Fe/H] value provided in {\it Gaia} DR3 (Fig.~\ref{fig:feh}). The LSP stars show higher metallicity than their non-LSP counterparts. 

\begin{figure*}
   \centering
       \includegraphics[width=0.85
       \textwidth]{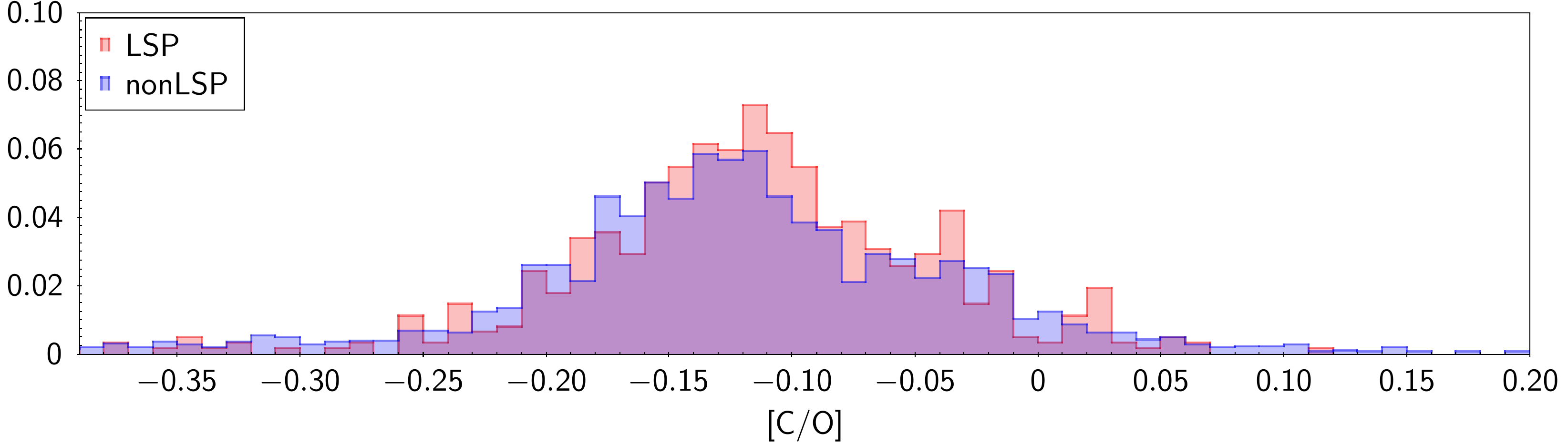} 
 \caption{Distribution of the C/O ratio based on the APOGEE data. The level of statistical significance tested with KS test is $p$ = 0.00083.}
 \label{fig:co}
\end{figure*}

Next, we check the C/O ratio in the APOGEE data, which is the only of the three aforementioned surveys that has enough C and O abundance data available to make a statistically significant comparison. The comparison of the C/O distribution for LSP and non-LSP stars is shown in Fig.~\ref{fig:co}. The LSP stars appear to be more C-rich than the non-LSP ones. 

\begin{figure*}
   \centering
       \includegraphics[width=0.86\textwidth]{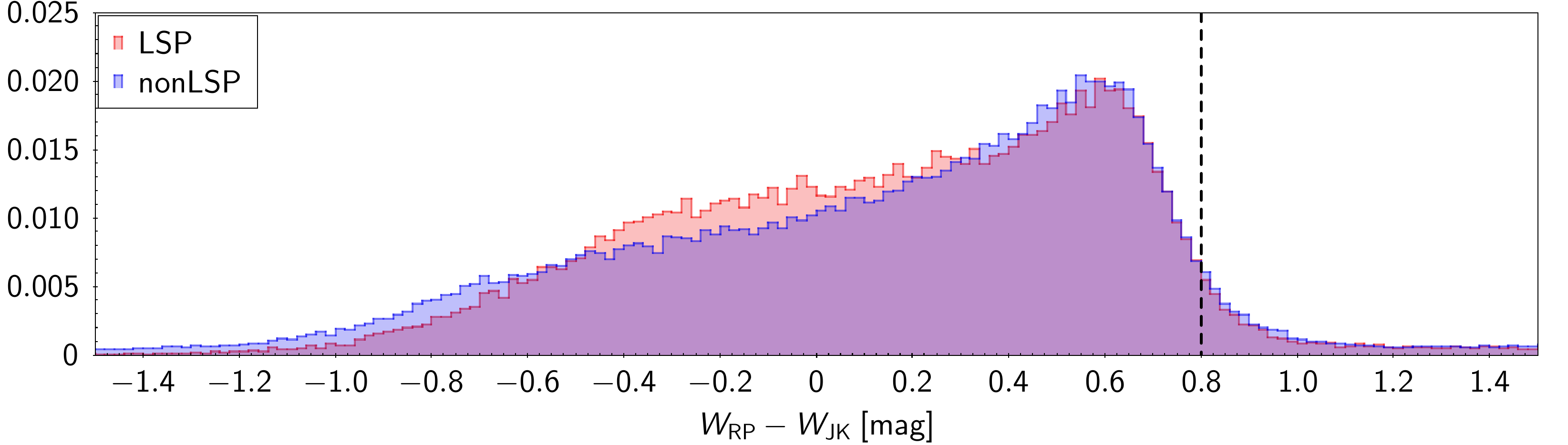} 
 \caption{$W_{RP}-W_{JK}$ {\it Gaia}-2MASS Wesenheit index, which can be used as an indicator C/O status. The level of statistical significance tested with the KS test is $p = 1.55\times 10^{-8}$. The vertical, dashed line at $W_{RP}-W_{JK}=0.8$ marks the boundry between O- and C-rich stars.}
 \label{fig:w}
\end{figure*}

\begin{figure*}
   \centering
       \includegraphics[width=0.85\textwidth]{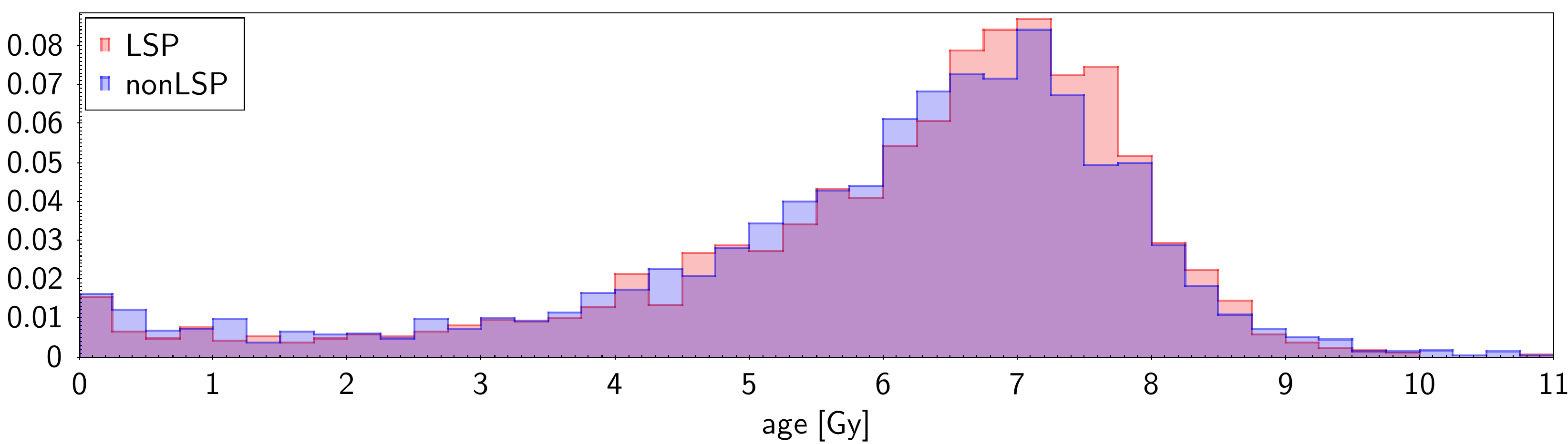} 
 \caption{Age distribution based on the  Rave data. The level of statistical significance tested with the KS test is $p$ = 0.043.}
 \label{fig:age}
\end{figure*}

We also use  the information about the C/O ratio provided in the {\it Gaia} DR3 catalog of Long Period Variables \citep{lebzelter2022}, based on {\it Gaia} spectro-photometric data. The fraction of C-rich stars in the LSP and non-LSP sample is comparable and in general low. In contrast with the APOGEE data, the {\it Gaia} chemical classification indicates a slightly lower fraction of C-rich stars of 4\% compared to 5\% in the non-LSP LPVs. 

To further investigate the C/O ratio of the stars in the sample we use the {\it Gaia}-2MASS Wesenheit index $W_{RP} -W_{JK}$ adopted from \citep{lebzelter2019}. This quantity is an indicator of the C/O, as it tends to be larger than 0.8~mag for C-rich and smaller than 0.8~mag for O-rich stars. The distribution in $W_{RP} -W_{JK}$ index is shown in Fig.~\ref{fig:w}. While the difference in the two distributions seems to be small, it turns out to be statistically significant when tested with the Kolmogorov-Smirnov (KS) test.

Another potentially interesting feature can be observed in the stellar age distribution derived from the RAVE data (Fig.~\ref{fig:age}). The LSP stars appear to be on average older than the non-LSP stars. The difference in age is small but still statistically significant at a 4\% level when tested with the KS test. 

\subsection{Quality Check}

As a first quality check, we repeat the analysis replacing the original comparison sample with the randomized one, and compare the results. The distribution in the randomized control sample is consistent with the base control sample, leading to the conclusion that the construction of the control sample does not introduce a significant bias. 

Another potential source of bias is the footprint of the ASAS-SN survey which is clearly visible in Fig.~\ref{fig:sky_all}. It results from the fact that the ASAS-SN catalog of variable stars is not homogeneous, as it includes both variable stars discovered independently by ASAS-SN, as well as stars from the literature catalogs. The most prominent feature of this footprint is the artificial over-density stripe around the galactic equator. In order to make sure that this artifact is not affecting our conclusions, we redo the analysis excluding the stripe between $-5^{\circ} < b < 5^{\circ}$, where the artifact appears. We obtain results consistent with the original analysis.

\begin{figure}
   \centering
       \includegraphics[width=0.45\textwidth]{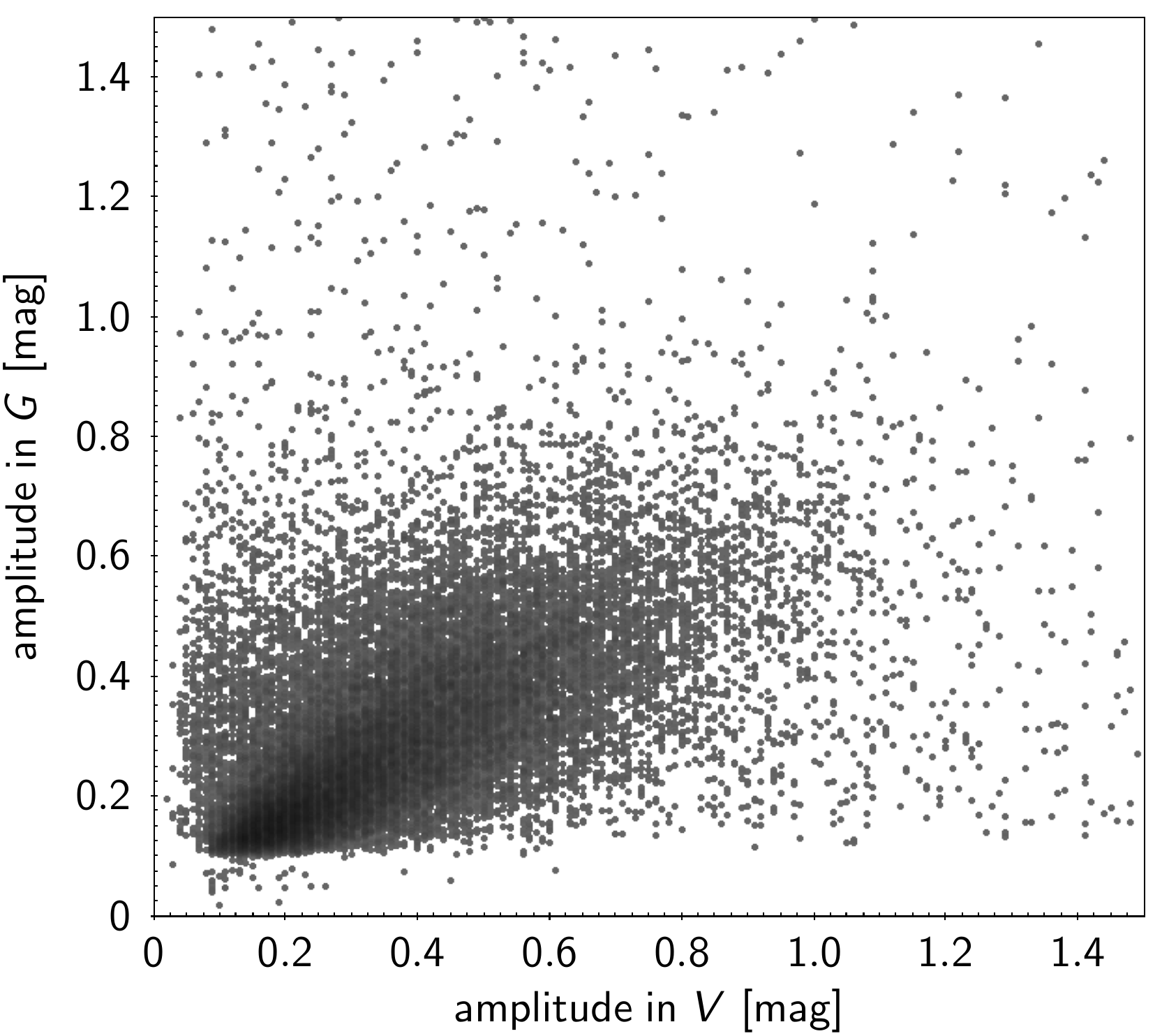} 
              \includegraphics[width=0.45\textwidth]{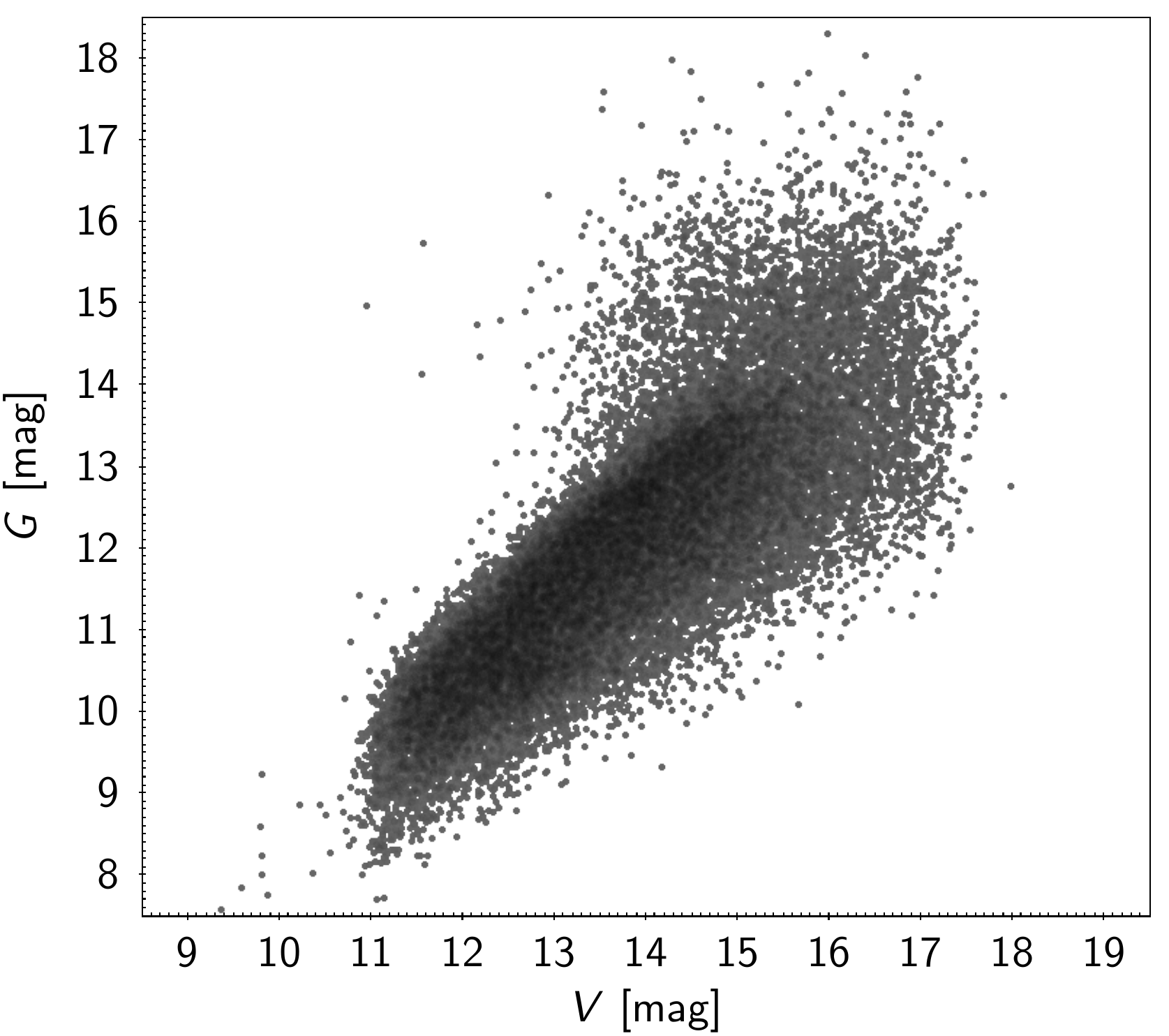} 
 \caption{ASAS-SN $V$-band amplitude vs. {\it Gaia} $G$-band amplitude (top panel) and mean $V$-band magnitude vs. mean  $G$-band magnitude (bottom panel). }
 \label{fig:ag_av}
\end{figure}

We note that, in crowded regions, the flux from unresolved stars can cause the photometric amplitude of a variable star to be underestimated \citep{riess2020}. \citet{lebzelter2022} have shown this to be the case for the $V$-band amplitude of ASAS-SN sources towards the Galactic plane and bulge. As we identify LSPs by their amplitude, this effect could lead us to mistake the pulsation period of a Mira in a crowded sky are as an LSP. In order to assess the impact of this effect, we compare the ASAS-SN $V$-band amplitude with the $G$-band amplitude from the 2$^{\rm nd}$ {\it Gaia} Catalog of Long Period Variable Candidates \citep{lebzelter2022}. The latter value is expected to be more accurate owing to the higher spatial resolution achieved by {\it Gaia}, and is available for about $20\:000$ objects from our sample. We find that, in most cases, the two amplitudes as well as the two magnitudes appear to be related by a power-law with some scatter (Fig.~\ref{fig:ag_av}). About 4\% of the sources deviate from this pattern and have a rather small $V$-band amplitude, indicative of the crowding effect. We repeated our analysis after excluding these sources and found the results to be consistent with the original analysis.

The last check that we make is related to the time coverage of the ASAS-SN photometry. While most of the light curves cover more than 1000 days of observation, there is a subset with shorter time coverage. For these objects, the LSPs may not always be reliably detected. As a final quality check, we exclude those stars with less than 1000 days of data and once more redo the analysis. Once again, the results are consistent. We conclude that the aforementioned factors do not introduce a significant bias to our analysis. 

\section{Discussion and Conclusions}

Our first observation is that most of the LSP stars lie on the first-overtone sequences B and C'. While there is a number of LSP stars that have one of the periods detected in the region roughly consistent with the fundamental mode sequence C, we conclude that these are most likely harmonics of the LSP, not real fundamental mode pulsations. 

The fact that the LSP stars appear predominantly on the first-overtone sequences B and C' is consistent with the previous results of \citet{trabucchi2017}, connecting the LSP to these two sequences and the transition between them. This result suggests that there may indeed be a physical reason tying the LSP phenomenon to the first overtone sequences, but the exact nature of this mechanism remains unclear. 

However, it should also be noted that the whole ASAS-SN sample of the LPVs, that we used for this search, is dominated by first-overtone pulsators. The fact that there are much fewer fundamental mode than overtone pulsators in the base sample may affect the distribution of the selected LSPs. Therefore, this result should be interpreted with caution.
 
The comparison between the LSP and non-LSP LPVs reveals some subtle, but statistically significant differences between the two populations. First, there is a difference between the spatial distribution of the two samples. The LSP stars appear to be more dispersed and less concentrated around the Galactic plane than their non-LSP counterparts, which hints at a possible age difference. 
 
The spectroscopically derived ages from the RAVE catalog seem to support the age difference hypothesis, as the LSP stars appear to be statistically older. It should be noted that the RAVE age estimates carry a high level of uncertainty and the statistical difference between the two distributions is marginally significant, therefore these results need to be interpreted with caution. Being aware of these limitations, the spectroscopic ages may be another hint of a population difference between LSP and non-LSP giants.
 
Another potentially important feature distinguishing the LSPs from non-LSPs is the C/O ratio, which has a statistically different distribution for the two samples. This can be seen both in the photometric data ($W_{RP} -W_{JK}$,  {\it Gaia}-2MASS Wesenheit index) and the specto-photometric classification provided in {\it Gaia} DR3, as well as direct spectroscopic measurements. Interestingly, the APOGEE data show a higher C/O ratio for the LSP stars, while the {\it Gaia} data seem to show a trend that is harder to interpret. While a statistically significant difference between the two populations can be observed, the direction of the trend is ambiguous. It should be noted that the APOGEE data is only available for a relatively small number of stars, located in very specific sky regions, mostly towards the Galactic Bulge, which may affect the result. On the contrary, the {\it Gaia} data is all-sky and therefore more representative of the general picture. 

The difference in C/O ratio is interesting for two reasons. First, the C/O depends on the metallicity \citep{bensby2006, marigo2020}. Therefore, it can be used as an indirect population indicator. The most obvious choice for such an indicator would be [Fe/H] itself. However, the number of objects that have reliable [Fe/H] values in any of the three spectroscopic catalogs used for this study is too small to draw a significant conclusion. Still, the difference in C/O can be interpreted as a hint towards the aforementioned hypothesis on the population difference. A the same time, the [Fe/H] values provided in {\it Gaia} DR3 indeed show different distributions, with LSP having higher metallicity on average. 
 
Second, the difference in C/O ratio may also be connected to the high mass-loss rate, as dust-driven wind requires a certain level of carbon abundance to be activated \citep{lagadec2008,goldman2017,marigo2020}. The mass loss and dust production are, in turn, known to be connected to the LSP phenomenon \citep{mcdonald2019}. The dust production becomes especially important in the case of the binarity scenario, which assumes a presence of a dust cloud around a low-mass companion \citep{soszynski2021}. Therefore, the difference in terms of C/O ratio between the LSP and non-LSP giants may be seen as an indicator of the binarity hypothesis. 

However, the potential population difference are rather subtle, making it hard to draw definitive conclusions. Further studies are required to verify this hypothesis. With coming improved spectroscopic data the type diagnostic we do will become more stringent, or will be able to confirm the tendencies we see.

\begin{acknowledgements}
MP is supported by the SONATINA grant 2020/36/C/ST9/00103 from the Polish National Science Center and the BEKKER fellowship BPN/BEK/2022/1/00106 from the Polish National Agency for Academic Exchange. LE and MP acknowledge the funding of the Department of Astronomy of the University of Geneva. MT and NM acknowledge the support provided by the Swiss National Science Foundation through grant Nr. 188697.

This work made extensive use of TOPCAT \citep{taylor2005} and VARTOOLS \citep{hartman2016}. 

This work has made use of data from the European Space Agency (ESA) mission {\it Gaia} (\url{https://www.cosmos.esa.int/gaia}), processed by the {\it Gaia} Data Processing and Analysis Consortium (DPAC, \url{https://www.cosmos.esa.int/web/gaia/dpac/consortium}). Funding for the DPAC has been provided by national institutions, in particular the institutions participating in the {\it Gaia} Multilateral Agreement.

\end{acknowledgements}


\end{document}